\documentclass[12pt,preprint]{aastex}

\begin{document}

\title{A NEW TECHNIQUE FOR HETERODYNE SPECTROSCOPY: LEAST-SQUARES
FREQUENCY SWITCHING (LSFS)}

\title {\today}

\author{Carl Heiles}

\begin{abstract}
We describe a new technique for heterodyne spectroscopy, which we call
{\it Least-Squares Frequency Switching}, or {\it LSFS}. This technique
avoids the need for a traditional reference spectrum, which---when
combined with the on-source spectrum---introduces both noise and
systematic artifacts such as ``baseline wiggles''. In contrast, LSFS
derives the spectrum directly, and in addition the instrumental gain
profile. The resulting spectrum retains nearly the full theoretical
sensitivity and introduces no systematic artifacts.

Here we discuss mathematical details of the technique and use numerical
experiments to explore optimum observing schemas. We outline a
modification suitable for computationally difficult cases as the number
of spectral channels grows beyond several thousand. We illustrate the
method with three real-life examples. In one of practical interest, we
created a large contiguous bandwidth aligning three smaller bandwidths
end-to-end; radio astronomers are often faced with the need for a larger
contiguous bandwidth than is provided with the available correlator.
\end{abstract}


\section{INTRODUCTION}  \label{introduction}

In digital heterodyne spectroscopy, the measured spectrum is the product
of the radio-frequency (RF) power and the intermediate-frequency (IF)
gain spectra; to obtain the RF power spectrum, one must divide the
on-source measured spectrum (the ON spectrum) by the IF gain
spectrum. This is usually accomplished by dividing by a reference
spectrum (the OFF spectrum), which is obtained by moving off in
frequency or position. However, using such OFF spectra introduces
additional noise because some observing time is spent off-source, and
also introduces additional artifacts (``baseline wiggles'').

In particular, obtaining accurate spectral profiles for Galactic
HI is difficult because of the difficulty in obtaining a good OFF
spectrum. There is no place in the sky where Galactic HI does not exist,
so one cannot use an OFF position.  Instead, the OFF spectrum is
commonly obtained by taking an off-frequency spectrum, moving far enough
in frequency so that the HI line is zero. This technique is known as
``frequency switching''.

	At many telescopes, frequency switching produces inaccurate OFF
spectra. This occurs because the RF gain and/or the RF power have
frequency structure. One contributor is reflections on the telescope
structure, for example between the feed and the reflector. If their
separation distance is $D$, then the reflected signal returns with a
time delay $\tau = {2D \over c}$. This produces a peak in the
autocorrelation function of the received signal with delay $\tau$, which
in turn produces a sinusoidal ripple in the frequency spectrum with
period $f_\tau = {1 \over \tau}$. At Arecibo, $f_\tau \sim 1.0$ MHz,
equivalent to about 200 km s$^{-1}$, which is comparable to the velocity
ranges of interest for many HI studies.  Similarly, mm-wave telescopes
used for molecular emission are much smaller than Arecibo and the line
frequencies are much larger, and again the ripple is comparable to
interesting line widths. Telescopes typically have many reflecting paths
with different delays, so the received signal has a superposition of
ripples with somewhat different periods. These ripples cannot be removed
by frequency switching, and in fact are sometimes amplified by an
unfortunate choice for the frequency-switching interval.

	Here we describe a new approach. Instead of switching the local
oscillator (LO) frequency and hoping for good cancellation of the
ripple, we set the LO frequency to a number $N$ of different values so
that we can {\it evaluate} the RF power spectrum and the IF gain
spectrum as distinct entities using a least-squares technique; we
call this {\it Least-Squares Frequency Switching}, or {\it
LSFS}. We begin in \S \ref{review} by reviewing the conventional
switching techniques; these introduce extra noise and baseline
artifacts, both of which are reduced or eliminated by LSFS.

The rest of the paper is devoted to LSFS. \S \ref{lsfs0} describes the
basics of the technique. \S \ref{realworld} illustrates our first
observational attempt, in which we created a large contiguous bandwidth
by aligning three smaller bandwidths end-to-end. The method relies on
choosing sensible LO frequencies, and with unwise choices the
least-squares matrices can be degenerate; \S \ref{details} discusses
this problem and its solution using Singular Value Decomposition.  \S
\ref{schemas} presents several schemas for choosing LO frequencies, and
\S \ref{numexpts} presents the results of numerical experiments that
evaluate the quality of these schemas. Up to this point, all of the
discussion is directed towards total power in a single polarization, or
alternatively Stokes $I$; \S \ref{stokes} shows how the technique
applies to the polarized Stokes parameters $(Q,U,V)$. Finally, \S
\ref{comparison} compares switching with LSFS and \S \ref{summary} is a
summary.

\section{REVIEW OF FUNDAMENTALS: POSITION AND FREQUENCY SWITCHING}
\label{review}

	In heterodyne spectroscopy, we convert the radio-frequency (RF)
spectrum to a low intermediate frequency (IF). This conversion is done
by multiplying RF power by the local oscillator (LO) in the mixer.
Symbolically, denoting frequency by $f$, we multiply $f_{RF}$ by
$f_{LO}$. This multiplication generates the sum and difference
frequencies $(f_{RF} + f_{LO})$ and $f_{IF}=(f_{RF} - f_{LO})$; we
remove the sum with a suitable low-pass filter, leaving the desired
near-baseband $f_{IF}$. 

	The mixer is a transition point between RF and IF frequencies.
We can meaningfully discuss the RF and IF sections as separate entities.
Thus, the RF section receives from the sky the antenna temperature
$T_A(f_{RF})$ and also has the receiver contribution $T_R(f_{RF})$,
which is often much larger. These are multiplied by the RF transfer
function, known as the gain $G_{RF}(f_{RF})$. Most of the frequency
dependence of $G_{RF}(f_{RF})$ occurs in the feed and electronics, which
operates on both $T_A$ and $T_R$, at least to a first approximation.
However, often the antenna temperature suffers an additional
frequency-dependent gain, which occurs because the incoming power
reflects from various portions of the telescope structure and interferes
with itself. For simplicity, we neglect this difference and assume that
the RF gain affects $T_A$ and $T_R$ equally. Thus, symbolically, the RF
power into the mixer $S_{RF}(f_{RF})$ is equal to

\begin{equation} \label{two}
S_{RF}(f_{RF}) = G_{RF}(f_{RF}) \ [T_A(f_{RF}) +T_R(f_{RF})] \ .
\end{equation}

	The IF section has a transfer function, or gain,
$G_{IF}(f_{IF})$. A well-designed system has no additional power
contributed at IF.  The spectral power measured by the digital
spectrometer is provided as a function of the IF frequency, so the
appropriate symbol is $P_{IF}(f_{IF})$, which is given by

\begin{equation} \label{measuredpower}
P_{IF}(f_{IF}) = G_{IF}(f_{IF}) \ S_{RF} (f_{RF}) \ .
\end{equation}

\noindent The relationship between the
spectral channels and IF frequency is fixed. We access different
portions of the RF spectrum by changing the LO frequency. 

	We now break the $T_A(f_{RF})$ and $T_R(f_{RF})$ into
frequency-independent (``continuum'') and frequency-dependent
(``spectral'') portions to simplify further development.  With these
decompositions, the measured spectral power $P_{IF}(f_{IF})$ depends on
the following quantities: \begin{enumerate}

\item $T_A(f_{RF})$.  The subscript $A$ means ``antenna temperature'',
so this is the spectral line contribution from the {\it sky}. The
explicit presence of the dependence $(f_{RF})$ means that there is a
spectral dependence, as befits a spectral line or the usually
slowly-varying continuum radiation. 

\item $T_A$, the frequency-independent portion of the
antenna-temperature continuum contribution. Continuum radiation is
weakly dependent on frequency; the frequency-dependent portion is
incorporated into $T_A(RF)$. 

\item $T_R(f_{RF})$. The subscript $R$ means ``receiver temperature'',
and includes all non-antenna contributions. As with $T_A(f_{RF})$, the
explicit dependence on $(f_{RF})$ denotes only the frequency-dependent
portion. For many systems, spectral variations in $T_R(f_{RF})$ are
fractionally small. 

\item $T_R$, the receiver-temperature continuum (frequency-independent)
contribution. 

\item $G_{RF}(f_{RF})$, the RF gain (dependent only on RF frequency).
For many systems, $G_{RF}(f_{RF})$ varies slowly with frequency.

\item $G_{IF}(f_{IF})$, the IF gain (dependent on IF frequency). With
digital spectroscopy, we must limit the bandwidth by an appropriate
IF bandpass filter. This means that $G_{IF}$ varies severely across the
band, varying from 0 to full gain.

\end{enumerate}

	With these definitions equation \ref{two} becomes a bit more
elaborate. When we look at a source in the sky we measure the on-source
(ON) spectrum

\begin{equation} \label{basic}
P(f_{IF}) = G_{IF}(f_{IF}) \ G_{RF}(f_{RF}) \ 
        \left[ (T_A(f_{RF}) + T_A) + (T_R(f_{RF}) + T_R) \right] \ .
\end{equation}

\noindent Our goal is to disentangle the sky contribution from
everything else, i.e.\ to obtain $(T_A(f_{RF}) +T_A)$.  Being primarily
interested in spectroscopy, a modified goal is to obtain only the
spectral portion $T_A(f_{RF})$.

	We cannot do either of these without dealing with the two gains and
the contributions from $T_R$. It is traditional to deal with these
extraneous quantities by taking a reference spectrum, usually denoted
the off-source (OFF) spectrum, and arithmetically combining it with the
ON spectrum by taking $\left(ON - OFF \over OFF\right)$. This process is
commonly known as ``switching''. It works well if the frequency
dependencies of the extraneous quantities are benign, but this is not
always the case. Let us examine the results of this switching process.

	Below we consider two ways of obtaining the OFF, one by moving
off in position and one by moving off in frequency. Let {\it primed}
quantities indicate the OFF measurements and {\it unprimed} the ON.
Further, let us simplify the problem by assuming the spectral dependence
of the receiver temperature to be small, i.e. 

\begin{equation} \label{assumption1}
T_R(f_{RF}) \ll T_R \ .
\end{equation}

\noindent This allows us to make a Taylor expansions of the expression
$\left(ON - OFF \over OFF\right)$ for the two types of switching by
dropping terms higher than first order. (We make these expansions to
minimize complication; if higher-order terms are included, the
expressions become more complicated but the techniques can be still
applied). 

\subsection{Position Switching}

	When the astronomical source is limited in angular extent we can
obtain the OFF spectrum by pointing the telescope away from the source.
For simplicity, we further assume that the OFF position has no line.
Remembering that $T_R=T_R'$, this gives for ${ON - OFF \over OFF}$

\begin{equation} \label{posswitch}
{P(f_{IF}) -P'(f_{IF}) \over P'(f_{IF})} = 
	[T_A(f_{RF}) + (T_A - T'_A)] 
\left[{1 - {T_R(f_{RF}) \over (T'_A + T_R)}  \over T'_A + T_R }\right] \ .
\end{equation}

	This gives the desired quantity $T_A(f_{RF})$ plus the additive
constant $(T_A-T'_A)$, which is the difference between the antenna
temperatures of the two positions. The result is further contaminated by
the right-hand multiplicative factor. In effect, this is a
frequency-dependent gain. However, its effect on the line shape is small
because of our assumption of equation \ref{assumption1}. These small
effects mean that position switching is usually the technique of choice.

	However, this does not mean that position switching always
provides good results. If the difference in continuum temperatures $(T_A
- T'_A)$ is large, then its multiplication by the right-hand
multiplicative factor produces a large effect, and this can make it
impossible to distinguish the astronomical spectral line $T_A(f_{RF})$. 
Thus, position switching can fail for weak lines with strong continuum
sources.

\subsection{Frequency Switching}

	If the spectral line is sufficiently spatially extended then we
cannot position switch. The prime example is Galactic HI. Here one
normally moves off in frequency. This means that the ON and OFF RF
frequencies differ, i.e.\ $f_{RF} \neq f'_{RF}$. In particular, the RF
gains differ between the ON and OFF measurements; also, the continuum
antenna and receiver temperatures subtract out, i.e.\ $T_A=T_A'$ and
$T_R = T_R'$. Again, to simplify, we assume the ON (unprimed) and OFF
(primed) gains do not differ much, i.e.\ we define 

\begin{mathletters}
\begin{equation}
{\Delta G \over G} = 1 - {G'_{RF}(f_{RF}) \over G_{RF}(f_{RF})}
\end{equation}

\noindent and assume 

\begin{equation}
{\Delta G \over G} \ll 1 \ .
\end{equation}
\end{mathletters}

\noindent The differing gains introduce a further complication into the
Taylor expansion: 

\begin{equation} \label{freqswitch}
{P(f_{IF}) -P'(f_{IF}) \over P'(f_{IF})} = 
\left[ T_A(f_{RF}) + (T_R(f_{RF}) - T'_R(f_{RF}))+ 
{\Delta G \over G}(T_A + T_R + T_A(f_{RF})) \right] 
\left[{1 - {T_R(f_{RF}) \over (T_A + T_R)}  \over T_A + T_R }\right] \ .
\end{equation}

\noindent This is similar to equation \ref{posswitch} except for the
additive term ${\Delta G \over G}(T_A + T_R + T_A(f_{RF}))$ in the first
factor on the right-hand side. Even though ${\Delta G \over G} \ll 1$,
this term is disastrous because it operates on $T_R$, which is large.
Unless ${\Delta G \over G} \lll 1$, this combination produces serious
baseline contamination in frequency switching. Nevertheless, frequency
switching works well when, as is often the case, $\Delta G \over G$
varies smoothly and slowly with $f_{RF}$ so that it is well-represented
by a low-order polynomial fit.

{\boldmath
\section{DETERMINATION OF $G_{IF} (f_{IF} \! )$ BY LEAST-SQUARES FREQUENCY
SWITCHING (LSFS)} \label{lsfs0}
}

	The classical approaches of position and frequency switching
work only under good conditions. The quantity having the most severe
frequency variations is $G_{IF}(f_{IF})$. If we could determine this
quantity explicitly we could forgo the switching and, instead, simply
divide all measured spectra $P(f_{IF})$ by $G_{IF}(f_{IF})$.
Least-squares frequency switching (LSFS) does this explicit
determination. 

\subsection{The Basic Equations and their Iterative Solution}

\label{iterative}

	We begin by rewriting equation \ref{basic} in a much simpler
form. Its right-hand side is the product of the IF gain and several
RF quantities. We lump these RF quantities into a single one,
$S_{RF}(f_{RF})$: 

\begin{equation} \label{define_s}
S_{RF}(f_{RF}) = G_{RF}(f_{RF}) \ 
        \left[ (T_A(f_{RF}) + T_A) + (T_R(f_{RF}) + T_R) \right] 
\end{equation}

\noindent so we rewrite equation \ref{basic} as

\begin{equation} \label{basic1}
P(f_{IF}) = G_{IF}(f_{IF}) \ S_{RF}(f_{RF}) \ .
\end{equation}

\noindent Our technique extracts $G_{IF}(f_{IF})$ and $S_{RF}(f_{RF})$
as separate entities.

	To proceed, we first express frequencies as channel offsets. The
digital spectrometer produces a spectrum having $I$ channels, with
channel number $i$ ranging from $i=0$ to $i=I-1$. The frequency
separation between adjacent channels is $\Delta f$. Thus, for the IF
frequency of channel $i$ we can write 

\begin{equation}
f_{IF,i} = f_0 + i \Delta f \ ,
\end{equation}

\noindent where $f_0$ is a constant. The separation $\Delta f$ also
applies to $f_{RF}$, so apart from a possible additive constant we have
for the RF frequency of channel $i$

\begin{equation}
f_{RF,i} = f_0 + f_{LO} + i \Delta f \ ,
\end{equation}

\noindent where $f_{LO}$ is the LO frequency. 

	In LSFS, we make measurements at $N$ different LO
frequencies, each designated by $n$. We increment these frequencies in
units of $\Delta f$, and we write

\begin{equation}
f_{LO,n} = f_{LO,n=0} + \Delta i_n \Delta f \ ,
\end{equation}

\noindent where $n$ ranges from 0 to $N-1$. $\Delta i_n$ is the number
of channels that $f_{LO,n}$ is offset from $f_{LO,n=0}$; clearly,
$\Delta i_{n=0}=0$. For convenience we assume $\Delta i_n$ to increase
monotonically with $n$, so the maximum LO excursion is $\Delta i_{N-1}$.
We can write all frequencies in units of the channel separation $\Delta
f$, so the RF frequencies become expressed as digital indices $i +
\Delta i_n$, where $i$ is the IF frequency offset from spectral channel
zero and $\Delta i_n$ is the LO frequency offset from the lowest LO
frequency (at $n=0$), both in units of the channel width $\Delta f$.
Equation \ref{basic1} becomes

\begin{equation} \label{lsone}
P_{i, \Delta i_n} = G_i S_{i + \Delta i_n} \ .
\end{equation}

\noindent \S \ref{textbook} presents the simplest ``textbook'' example, including
a figure, to help explain the somewhat confusing relationships embodied
in the above description.

	There are $NI$ of these equations. We could use them as our
equations of condition for the least-squares fit. However, for reasons
discussed below, we normalize the variables for computational
efficiency. To normalize, we require the means over $(i,n)$, denoted by
$\langle P_{i, \Delta i_n} \rangle$ and $\langle S_{i + \Delta i_n}
\rangle$, to equal unity; of course, this implies that the typical $G_i
\sim 1$. Henceforth we assume $P$ and $S$ to be so normalized.

	In addition, we express $S_{i + \Delta i_n}$ (whose mean is
unity) as an offset $s_{i + \Delta i_n}$ from unity,  i.e.\ we write

\begin{equation}
S_{i + \Delta i_n} =  1 + s_{i + \Delta i_n} \ .
\end{equation}

\noindent Clearly, the mean $\langle s_{i + \Delta i_n} \rangle=0$.
Below we will assume $s \ll 1$, which should be valid as long as the
total fractional bandwidth ${[(I-1) + \Delta i_{N-1}] \Delta f \over
\langle f_{RF} \rangle}$ is not too large and, also, there are no strong
spectral lines. This assumption will be made only for reasons of
computational efficiency and does not affect the final solution. For
$P_{i, \Delta i_n}$, we can replace the index $\Delta i_n$ by the index
$n$ to reduce clutter. Equation \ref{lsone} becomes

\begin{equation} \label{basic2}
P_{i, n} = G_i + G_i \ s_{i + \Delta i_n} \ .
\end{equation}

\noindent The $P_{i,n}$ are measured quantities and the $G_i$ and $s_{i
+ \Delta i_n}$ are unknowns to be determined by a least-squares fit.
With $I$ spectral channels, $N$ LO frequencies, and a maximum LO
excursion of $\Delta i_{N-1}$ frequency channels, the total number of
unknowns is $a=(2I + \Delta i_{N-1})$: there are $I$ unknown values of
$G$ and $(I + \Delta i_{N-1})$ values of $s$. For example, for the
calibration spectrum of Arecibo's GALFA spectrometer (Stanimirovi{\'c} et
al.\ 2006), we have 512 channels so $I=512$ and we use $\Delta i_{N-1} =
31$, so there are 1055 unknowns.  This is a substantial, but hardly
impossible, least squares problem. Its solution requires, in essence,
the inversion of a $1055 \times 1055$ matrix.

	We must solve this set of equations using nonlinear
least-squares techniques, which are required because both $G$ and $s$
are unknown. Nonlinear least-squares is an iterative process, involving
making a guess for the parameters and solving for the difference between
the guesses and the true values. Let the guessed values of the
parameters be denoted by the superscript $g$. Then, taking the
difference between the true values and the guessed values, and using
equation \ref{basic2}, we have

\begin{equation}
P_{i,n}-P^g_{i,n} = (G_i + G_i \ s_{i + \Delta i_n})-
	(G^g_i + G^g_i \ s^g_{i + \Delta i_n})
\end{equation}

\noindent We express the differences between quantities with the symbol
$\delta$; thus the unknowns become $\delta G_i=G_i-G_i^g$ and $\delta
s_{i + \Delta i_n}= s_{i + \Delta i_n}-s^g_{i + \Delta i_n}$. As usual in
iterative schemes, we assume these differences are small and drop second
order terms. Also, we divide through by $G^g_i$, an act which implicitly
assumes that its fractional error is small, but as we shall see in fact
it does not matter if its fractional error isn't small. This gives

\begin{equation}
{\delta P_{i,n} \over G^g_i} = {\delta G_i \over G^g_i} (1 + s^g_{i +
\Delta i_n}) + \delta s_{i + \Delta i_n}
\end{equation}

\noindent We have turned the nonlinear least-squares problem into an
iterative linear one. 

	However, the presence of the term $ s^g_{i + \Delta i_n}$ means
that the equation-of-condition matrix changes from one iteration to the
next. The number of unknowns is large, and the need to evaluate the
inverse matrix for each iteration requires significant computational
time. We can eliminate this burden by dropping the term $s^g_{i + \Delta
i_n}$ in the above equation. This yields our final set of equations, in
which the two unknowns for each channel ($i$) and each LO setting
($\Delta i_n$) are now ${\delta G_i \over G^g_i}$ and $\delta s_{i +
\Delta i_n}$:

\begin{equation} \label{final_ls}
{\delta P_{i,n} \over G^g_i} = {\delta G_i \over G^g_i}	+ \delta s_{i +
\Delta i_n}
\end{equation}

\noindent We use this in an iterative solution in which each step is a
linear least-squares fit for the two sets of unknowns. The coefficients
of the unknowns are all equal to unity, so they remain constant from one
iteration to the next; in other words, the equation-of-condition matrix
in the least-squares treatment does
not change from one iteration to the next.

Now, it might seem that the elimination of the term  $s^g_{i +
\Delta i_n}$ is an arbitrary action that produces erroneous results.
However, this is not a problem. As ${\delta G_i \over G_i^g} \rightarrow
0$---i.e., as we attain convergence---this term goes to zero so the
final solution is unaffected. And, miraculously, it {\it does}
converge---usually rapidly. 

	The set of equations \ref{final_ls} does not sufficiently
constrain the solution. One more equation is needed to keep the mean
RF power $\langle S \rangle$ approximately constant (i.e.,
approximately equal to unity), which we include as an additional
equation of condition:

\begin{equation} \label{powerconservation}
\sum_{i,n} \delta s_{i + \Delta i_n} = 0
\end{equation}

	To solve this set of equations iteratively, we begin with
initial guesses $G^g_i = 1$ and $s^g_{i + \Delta i_n}= 0$. This provides
the initial $P^g_{i,n} = 1$. We least-squares solve the $NI$ equations
\ref{final_ls} and the equation \ref{powerconservation} for the
unknowns, which are $\delta G_i \over G^g_i$ and $\delta s_{i + \Delta
i_n}$. The solution provides the new guesses $G^g_i$ and $s^g_{i +
\Delta i_n}$. We usually obtain convergence in $\sim 10$ iterations,
which takes a fraction of a second on a contemporary laptop computer for
$NI=1055$.

One final comment. After the calculation is finished, the mean of the gains $\langle
G_i^g \rangle = {\cal G}$ ends up 
departing a bit from unity. For many purposes, e.g.\ when combining
independent LSFS results, it is desirable to scale
the gains so that their mean is unity. To accomplish this, simply divide
all the derived gains by ${\cal G}$, i.e.\ we write

\begin{mathletters}
\begin{equation}
G_{i, scaled} = {G_i^g \over {\cal G}} \ .
\end{equation}

\noindent Similarly, the RF powers are also scaled:
\begin{equation} 
S_{i + \Delta i_n, scaled} = {\cal G} \left( 1 + s^g_{i + \Delta i_n}
\right) \ .
\end{equation}
\end{mathletters}

\subsection{Number of Equations of Condition; Number of Unknowns}

\label{numbers}

	In least-squares fitting we develop equations of condition, one
for each observed quantity. Least-squares fitting requires the unknowns
to be overdetermined, i.e.\ the number of unknowns to be smaller than
the number of equations of condition.  As mentioned above just after
equation \ref{basic2}, the number of unknowns is $a = (2I + \Delta
i_{N-1})$. The number of equations of condition is $M= NI + 1$: $I$
channels for each of the $N$ LO frequencies, plus equation
\ref{powerconservation}. For the least-squares technique we require $M >
a$, i.e.

\begin{equation}
NI > 2I + \Delta i_{N-1} - 1 \ .
\end{equation}

\noindent Clearly, it makes no sense to have $\Delta i_{N-1} > I$,
because that generates additional values of $s_{i + \Delta i_n}$ while
providing no information on $G_i$. Thus we require $\Delta i_{N-1} = h
I$, where $h < 1$. This yields

\begin{equation} \label{nvsh}
N > 2 + h - {1 \over I}
\end{equation}

\noindent With $h < 1$, we generally require $N \geq 3$. The quantity
$h$ is maximum LO separation $\Delta i_{N-1}$ in units of the IF
bandwidth; we refer to $h$ as the {\it fractional LO coverage}.  

\subsection{An Illustrative Textbook Example} \label{textbook}

\begin{figure} [h!]
\begin{center}
\includegraphics[width=5.0in]{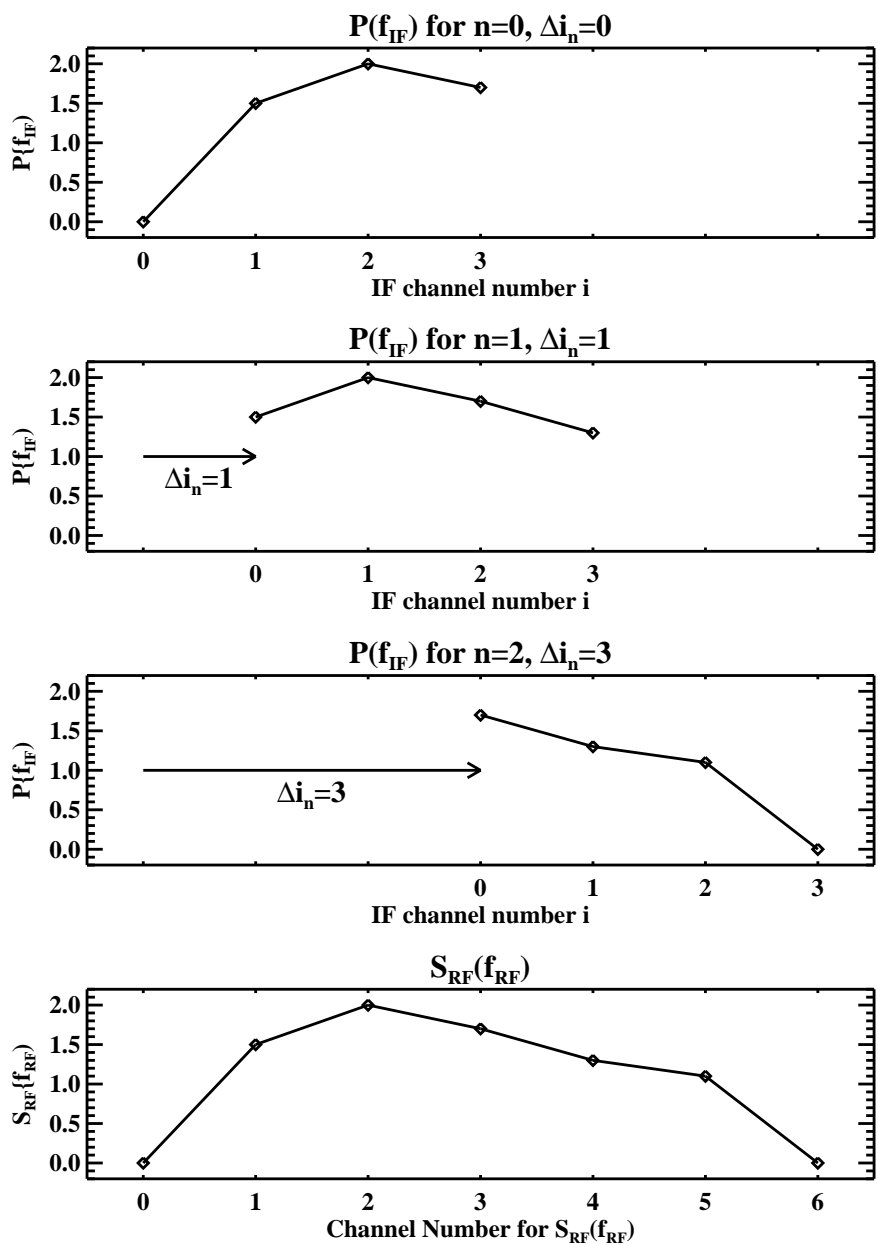} 
\end{center}
\caption{Graphical illustration of the illustrative textbook example,
  assuming all $G_{IF}(f_{IF}) = 1$. The top three panels show the
  measured IF power $P_{i, \Delta i_n}$ [which is captioned
  ``$P(f_{IF})$''] versus IF channel number $i$; the bottom panel shows
  the RF spectrum $S_{i+\Delta i_n}$ [which is captioned
  ``$S_{RF}(f_{RF})$''] versus $(i+\Delta i_n)$ (which is captioned
  ``Channel Number for $S_{RF}(f_{RF})$'').
  \label{xtrafig}}
\end{figure}

	We present an illustrative example with the goal of clarifying
the procedure. The simplest example has the smallest numbers. Since we
need $N \geq 3$ and $I > \Delta i_{N-1}$, we choose $I=4$ and $N=3$,
with $\Delta i_n = [0, 1, 3]$. That is, we have a 4-channel
spectrometer. We use 3 LO frequencies, with the latter two spaced by 1
and 3 channels from the first. This provides an arithmetic progression
for the successive frequency differences $\Delta i_{nn'}$, which we
define as $(\Delta i_n - \Delta i_{n'})$ (where we consider all
combinations of $n$ and $n'$). The values of $\Delta i_{nn'} = [0, 1, 2,
3]$. Note that $\max(\Delta i_{nn'}) = \Delta i_{N-1} = 3$. Figure
\ref{xtrafig} graphically illustrates these parameters; this Figure
assumes $G_{IF} (f_{IF})=1$ everywhere.

In matrix form, the equations of condition (equations \ref{final_ls} and
equation \ref{powerconservation}) are

\begin{equation} \label{matrixeq}
{\bf X \cdot a} = {\bf p} \ .
\end{equation}

\noindent Here, $\bf X$ has $(NI+1)$ rows and $(2I+\Delta i_{N-1})$
columns, $\bf a$ is the $(2I+\Delta i_{N-1})$ vector of unknowns, and $\bf
p$ is the $(NI+1)$ vector of $NI$ measured and one constrained
quantities.  Our notational convention is that boldface small letters
are vectors and boldface large letters are matrices.

For our textbook example, in the vector of unknowns $\bf a$, to avoid
clutter we write $g_i$ in place of ${\delta G_i \over G^g_i}$. To save
space, we write the transpose of this vector, which is

\begin{equation} {\bf a^T} = \left[ g_0, \;   g_1, \;   g_2, \;  
g_3, \;   \delta s_0, \;   \delta s_1, \;   \delta s_2, \;   
	\delta s_3, \;   \delta s_4, \;   \delta s_5, \;   \delta s_6 \right] \ .
\end{equation}

\noindent In the vector of measured quantities $\bf p$, we
write $p_{i,n}$ in place of $\delta P_{i,n} \over G^g_i$. This vector's transpose is

\begin{equation}
{\bf p^T} = \left[
p_{0,0} , \;  p_{1,0} , \;  p_{2,0} , \;  p_{3,0} , \;  
p_{0,1} , \;  p_{1,1} , \;  p_{2,1} , \;  p_{3,1} , \;  
p_{0,2} , \;  p_{1,2} , \;  p_{2,2} , \;  p_{3,2} , \;  0 
\right] \ ,
\end{equation}

\noindent and the equation-of-condition matrix consists of the
coefficients in equations \ref{final_ls} and \ref{powerconservation},
all of which are unity:
\begin{eqnarray} \label{xmatrixeqn}
{\bf X} = \left[
\begin{array} {ccccccccccc}
1 & 0 & 0 & 0 & 1 & 0 & 0 & 0 & 0 & 0 & 0 \\
0 & 1 & 0 & 0 & 0 & 1 & 0 & 0 & 0 & 0 & 0 \\
0 & 0 & 1 & 0 & 0 & 0 & 1 & 0 & 0 & 0 & 0 \\
0 & 0 & 0 & 1 & 0 & 0 & 0 & 1 & 0 & 0 & 0 \\
1 & 0 & 0 & 0 & 0 & 1 & 0 & 0 & 0 & 0 & 0 \\
0 & 1 & 0 & 0 & 0 & 0 & 1 & 0 & 0 & 0 & 0 \\
0 & 0 & 1 & 0 & 0 & 0 & 0 & 1 & 0 & 0 & 0 \\
0 & 0 & 0 & 1 & 0 & 0 & 0 & 0 & 1 & 0 & 0 \\
1 & 0 & 0 & 0 & 0 & 0 & 0 & 1 & 0 & 0 & 0 \\
0 & 1 & 0 & 0 & 0 & 0 & 0 & 0 & 1 & 0 & 0 \\
0 & 0 & 1 & 0 & 0 & 0 & 0 & 0 & 0 & 1 & 0 \\
0 & 0 & 0 & 1 & 0 & 0 & 0 & 0 & 0 & 0 & 1 \\
0 & 0 & 0 & 0 & 1 & 1 & 1 & 1 & 1 & 1 & 1 \\
\end{array}
\right] \ .
\end{eqnarray}

\noindent In this matrix, \begin{enumerate}

	\item The first four rows pertain to the lowest LO frequency with
$\Delta i_0 = 0$

	\item The next two pairs of four rows pertain to $\Delta i_1=1$
and $\Delta i_2= 3$.

	\item The last row is the power conservation equation
\ref{powerconservation}. 

	\item The first four columns are the coefficients of the four IF
	gains $\delta G_i \over G_i^g$ in equation \ref{final_ls}.

	\item The last seven columns are the coefficients of the seven
	RF powers $\delta s_{i + \Delta i_n}$ in equation \ref{final_ls}.
\end{enumerate}

\noindent The usual least-squares process of solving these equations of
condition (see Press et al.\ 1992) involves multiplying $\bf X$ by its
transpose to obtain the curvature matrix (the matrix of normal
equations) and then taking the inverse of that matrix product to obtain
the covariance matrix  {\boldmath $\alpha$}:

\begin{equation} \label{covariance}
\mbox{\boldmath$\alpha$} = {\bf (X^T \cdot X)^{-1}} \ ,
\end{equation}

\noindent and the solution for the coefficient vector is

\begin{equation} \label{stdsolution}
{\bf a} = \left( \mbox{\boldmath$\alpha$} {\bf \cdot X^T} \right) \bf{\cdot p} \ .
\end{equation}

\noindent For this illustrative problem, the inverse is well defined
with no numerical problems.  The biggest normalized covariance (or
correlation), obtained by converting the covariance matrix to a
correlation matrix, is $-0.51$, whose absolute value is not unreasonably
large. 

\section{AN ILLUSTRATIVE REAL-WORLD EXAMPLE} \label{realworld}

\begin{figure} [p!]
\begin{center}
\includegraphics[width=5.0in]{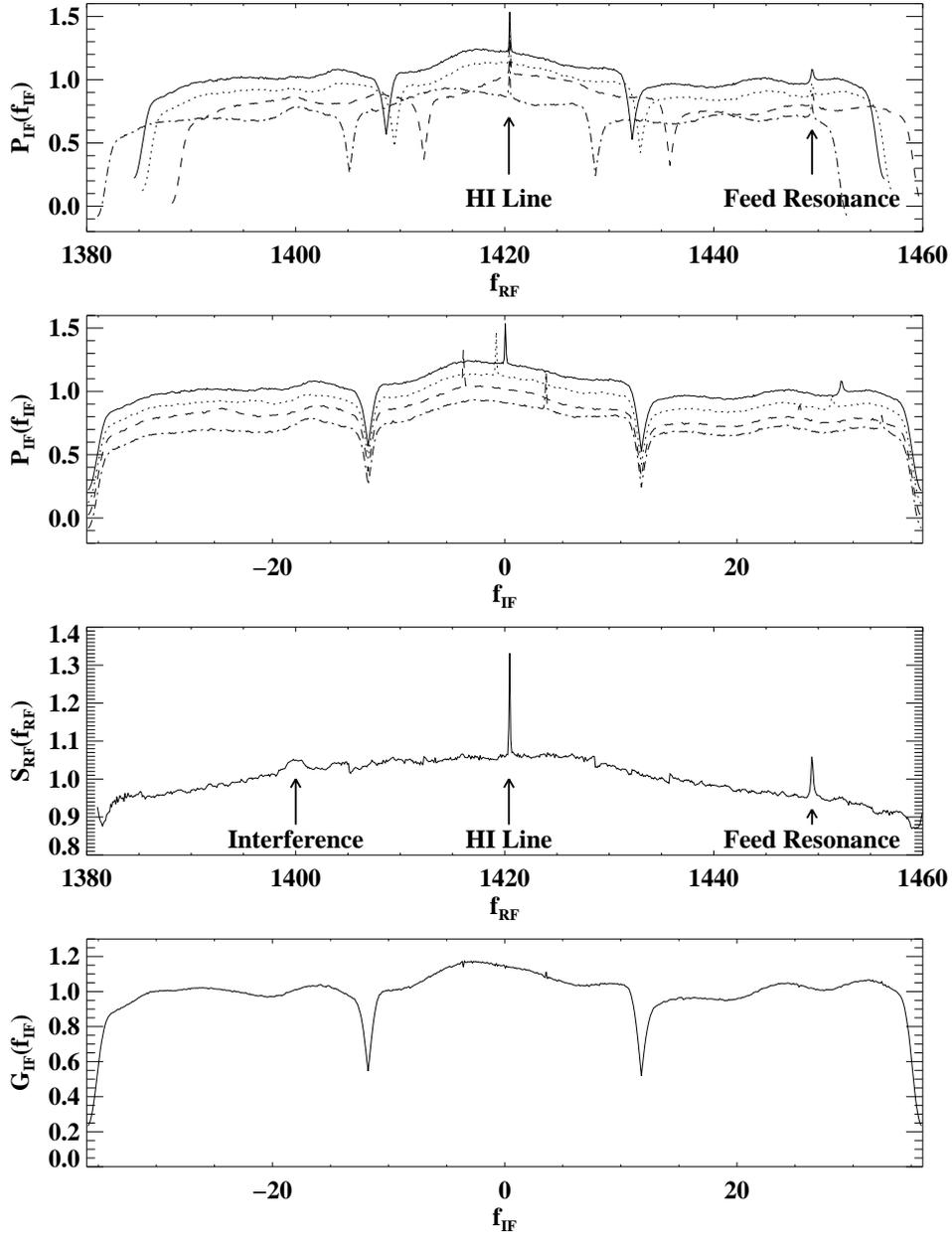} 
\end{center}
\caption{Real-world example of LSFS discussed in \S \ref{realworld}. The
top panel shows the total measured power $P_{IF}(f_{IF})$ versus RF
frequency for the four LO frequencies; note that the HI line remains
fixed near 1420 MHz and the IF bandpass shapes move with the LO. The
second panel shows the measured power $P_{IF}(f_{IF})$ versus IF
frequency for the four LO frequencies; the IF bandpass shapes remain
fixed and the HI line moves with the LO.  The third panel shows the
derived RF power, $S_{RF}(f_{RF})$ in equation \ref {basic1}, with the
known spectral peaks identified. The bottom panel shows the IF gain
$G_{IF}(f_{IF})$. \label{carl1plt}} 
\end{figure}

	Our initial experiments with LSFS were performed with three
independent banks of Arecibo's interim correlator\footnote{The
Arecibo Observatory is part of the National Astronomy and Ionosphere
Center, which is operated by Cornell University under a cooperative
agreement with the National Science Foundation.}. Each bank had 2048
channels and covered a bandwidth of 25 MHz; we overlapped the three
banks by cutting off 64 channels on each end, i.e.\ we spaced the
centers by ${31 \over 32} \times 25$ MHz, stitching together a single
spectrometer with 5888 channels covering a total contiguous bandwidth of
71.875 MHz. We binned the channels by a factor of 8, making 736 channels
of width 0.0977 MHz. We used four different LO frequencies with spacings
$\Delta i_n = [0, 35.50, 43.69, 72.36]$, providing the nonuniform set of
six spacings $\Delta i_{nn'} = [8.19, 28.67, 35.50, 36.86, 43.69,
72.36]$. Figure \ref{carl1plt} shows the results.

	LSFS works reasonably well in this initial-experiment
example. We performed this observation before devising the schemas
discussed in \S \ref{schemas} and we cannot remember how we chose the
set of LO frequencies $\Delta i_n$, but we suspect it is not a very good
choice because the middle third of the RF power spectrum $S_{RF}(f_{RF})$
is slightly displaced from the others. Such artifacts do not occur with
the better schemas of \S \ref{schemas}. It is clear that LSFS would
perform admirably with a better schema, and this particular case
illustrates how one can accomplish the often desirable goal of reliably
generating a contiguous large band from narrower ones.

We refer to two other real-world examples. In one, we use LSFS to
determine $G_{IF}(f_{IF})$ and use that gain spectrum to correct
thousands of measured spectra; we provide some details in \S
\ref{svdpractice}. In the other, we use LSFS to determine the intrinsic
ripples in the RF power spectrum; see \S \ref{baselinewiggles}.

\section{DETAILS OF MATRIX ALGEBRA FOR LARGER $I$, LARGER
$N$ } \label{details}

\subsection{Degeneracy in the {\bf X} Matrix}

	Under some conditions we find empirically that some of the
equations of condition become degenerate. This degeneracy is best
understood by considering the $\bf X$ matrix in equation \ref{matrixeq} as a
series of $a$ column vectors, where $a$ is the number of unknowns; $a =
2I + \Delta i_{N-1}$. Suppose that two columns are degenerate,
meaning that they are linear combinations of each other. In the matrix
product $\bf X \cdot a$, there is a one-to-one relationship between each
column in $\bf X$ and its corresponding unknown coefficient in $\bf a$. A
degeneracy between two columns means that the matrix product cannot
distinguish between the two corresponding coefficients.

	When one applies the usual technique of generating the normal
equations and inverting the curvature matrix, as in equation
\ref{covariance}, it does not work: the inverse matrix does not exist
because of the degeneracy. In cases like this one has two choices: think
hard and find the root cause of the degeneracies; or use Singular Value
Decomposition (SVD) to empirically remove them. The number of unknowns
is large, so the first option---picking through a huge matrix looking
for degeneracies---is difficult, probably even for a mathematical
expert. Therefore, we choose the latter one. 

\subsection{SVD: Theory} \label{svdtheory}

	Numerical Recipes (Press 1992) provides a useful discussion of
the SVD technique as applied to least squares. The SVD technique forgoes
the usual generation of the normal equations for calculating the matrix
$\bf \left( \mbox{\boldmath$\alpha$} \cdot X^T \right)$ in equation
\ref{stdsolution}. Rather, it expresses this matrix in terms of three
matrices that are derived from the SVD decomposition of the $\bf X$
matrix. 

	The cornerstone of SVD is that any $M \times a$ matrix, where
the number of rows $M$ and the number of columns $a$ satisfy $M \ge a$,
can be decomposed as the product of three matrices. In particular, our
matrix $\bf X$ in equation \ref{matrixeq} satisfies this criterion, so we
can write

\begin{equation} 
{\bf X} = {\bf U \cdot [W] \cdot V^T} \ ,
\end{equation}

\noindent where the right-hand side contains the three SVD
matrices. These matrices have important properties:
\begin{enumerate}

        \item $\bf U$ is $M \times a$, $\bf [W]$ is $a \times a$ and
diagonal, and $\bf V^T$ is $a \times a$. The square brackets around the
matrix $\bf W$ indicate that it is diagonal. 

        \item The columns of $\bf U$ consist of unit vectors that are
orthonormal, and the same is true for $\bf V$. Because $\bf V$ is
square, its rows are also orthonormal so that ${\bf V \cdot V^T} =
1$. Recall that, for square orthonormal vectors, the transpose equals
the inverse so $\bf V^T = V^{-1}$.
        
\end{enumerate}

	Degeneracies are directly reflected in the $\bf V$ and $\bf [W]$
matrices. The square matrix $\bf V$ consists of a set of $a$ orthonormal
column vectors. These are normalized orthogonal vectors that are linear
combinations of {\it non}orthogonal column vectors in $\bf X$.  Suppose
that $L$ columns of $\bf X$ are degenerate. Then the number of
independent orthonormal vectors represented in $\bf X$ is decremented by
$L$.  Nevertheless, the $\bf V$ matrix still contains $a$ independent
orthonormal vectors.  The decrement by $L$ is represented not by the
orthonormal column vectors in $\bf V$, but rather by their {\it weights}
in the diagonal matrix $\bf [W]$. Each column vector in $\bf V$ has an
associated weight in $\bf W$, and if there is degeneracy, then the
corresponding value of $\bf W$ is zero. This means the corresponding
orthonormal vector in $\bf V$ cannot be represented by the column
vectors in $\bf X$.

	Having derived the SVD components of $\bf X$, we can write for
the matrix product $\left( \mbox{\boldmath$\alpha$} {\bf \cdot X^T}
\right)$ in equation \ref{stdsolution}

\begin{equation} \label{svdcomps}
{\bf (\mbox{\boldmath$\alpha$} \cdot X^T)}  = 
{\bf V \cdot \left[ 1 \over W \right] \cdot U^T} \ .
\end{equation}

\noindent Now suppose there is degeneracy; then the corresponding values
of $\bf [W]$ are zero, so the corresponding values of $\bf \left[ {1 \over W}
\right]$ become infinite. This is an attempt by the matrix algebra to
represent the space defined by the corresponding columns of $\bf V$ with
data that were taken with inappropriate values of $\bf X$. 

	SVD, as applied to least squares, handles these infinities by
setting the corresponding values of $\bf [{1 \over W}]$ (which are formally
equal to $\infty$) to zero. This provides stable, realistic solutions in
which the offending degenerate coefficient values are close to being
correct---or, at least, not being totally unreasonable. By following
this procedure, one can handle degeneracies without understanding their
cause simply by zeroing out the relevant components of the $\bf \left[ {1
\over W} \right]$ vector.

	This zeroing process can---and should---be applied in cases of
{\it near} degeneracy. Just exactly what ``near'' means depends on the
noise in the data, because the data values are amplified by $\bf \left[ {1
\over W} \right]$ in calculating the coefficient values. For
sufficiently noisy data one might best zero out the offending elements
of $\bf \left[ {1 \over W} \right]$ even if they are not too terribly large. 

	The $\bf X$ matrix depends only on the values $\Delta i_n$, not
on the data values $S$. Moreover, calculating the inverse of the $\bf X$
matrix is computationally expensive. Thus, when using a particular set
of LO frequency offsets for multiple observations, it behooves one to do
the SVD calculation of ${\bf (\mbox{\boldmath$\alpha$} \cdot X^T)}$
once, store the results on disk, and read them back when necessary. This
has the further advantage that one can examine the weight vector $\bf
[W]$ once for each set of LO frequency offsets, decide which particular
values of $\bf \left[ 1 \over W \right]$ to set to zero, and forget
about dealing with this on a case-by-case basis. Section
\ref{svdpractice} provides some comments on examples we encountered.

\subsection{SVD: Practice} \label{svdpractice}

The ratio of maximum to minimum weights determines whether some inverse
weights should be zeroed. For noise-free data, ratios that come close to
the machine accuracy (perhaps $10^6$ for single-precision math) should
be zeroed; in the presence of noise, smaller ratios should be zeroed.
Our initial experiment of \S \ref{realworld} with four LO frequencies
had no degeneracy.  For our numerical experiments, we examined various
schemas, which are detailed in \S \ref{mra} and \S \ref{otherschemas}. Most
schemas had weight ratios smaller than 2500. The highest ratios occurred
for N=3 and decreased rapidly (e.g., by a factor of 2) for successive
increases in N. In some cases we zeroed inverse weights to keep the ratio
smaller than a few hundred, and in some cases not.

There were two exceptions, which had much larger ratios. The MRN,R
schemas had large ratios, again $\sim 10^7$, for $R \neq 1$. For $R=2$,
4, and 8 we had to zero 1, 3, and 7 inverse weights,
respectively. (Recall that this is out of a total of over 1000
weights). The $2^{MRN}$ schema had ratios $\sim 10^7$ for a few
elements; this schema is no good anyway and we do not quote results for
it here.

Our work with the GALFA spectrometer (Stanimirovi{\'c} et al.\ 2006) is
an interesting case.  This spectrometer observes two spectra
simultaneously, the wide ``calibration'' spectrum and the much narrower
``science'' spectrum.  The calibration spectrum is 100 MHz wide with 512
channels, for which we use the MR7 arrangement (see \S \ref{schemas}),
exactly like our MR7 numerical experiment below.

	We use the same set of 7 LO frequencies for the science
spectrum. This spectrum covers a bandwidth of ${ 100 \over 14} \approx
7.143$ MHz and has 8192 channels, which makes $\Delta f \approx 872$
Hz. Thus, the LO increment $\Delta f$ is about 224 times the channel
spacing. Of the 8192 channels, 7679 are recorded. We invent an extra,
making the total 7680. Before applying LSFS we rebin these, lumping
successive bins of 16 together so that the total number of binned
channels is 480. The frequency offset of the seventh LO frequency is
almost as large as the total bandwidth, so we use only the first six LO
frequencies. This gives increments $\Delta i_{nn'}$ ranging from 14 to
283, so $h= {283 \over 480} \approx 0.59$.  We assume smoothness and use
an interpolation procedure to recover the 7679 values of
$G_{IF}(f_{IF})$.

	This narrowband case exhibits thirteen degeneracies. We have
1198 coefficients to derive, so we have 1198 orthonormal column vectors
in the $\bf V$ matrix. The 1198 weights $\bf [W]$ range from about $8.9
\times 10^{-7}$ to 27. Thirteen weights are nearly degenerate, being
smaller than $1.8 \times 10^{-6}$; we set their corresponding inverses
to zero.  The lowest nonzeroed weight is $\approx 0.69$, so there is a
huge gap between the range of accepted weights ($\sim 0.69$ to 27) and
the zeroed inverse weights (the largest is $\sim 1.8 \times
10^{-6}$). Zeroing the 13 inverse weights provides a very nice
solution for the IF gain, which is used to correct thousands of
measured mapping spectra.

\section{REGARDING COMPUTING TIME} \label{comptime}
Regarding computing time, we scale from the MR7 scheme described in \S
\ref{schemas}, which has $I = 512$, $N=7$, and $\Delta i_{N-1}=31$.  The
LSFS $\bf X$ matrix {\it inversion} for $I=512$ and $\Delta i_{N-1}=31$
takes 123 seconds on a not-quite contemporary laptop computer programmed
in IDL. This time scales as the number of unknowns cubed, i.e.\ as $(2I
+ \Delta i_{N-1})^3$, so for $I=4096$ it would take about 20 hours.
This is a long time, but as discussed in \S \ref{lsfs0} and \S
\ref{details}, the matrix inversion should be done once and the result
stored on disk.  The $\bf X$ matrix is sparse, and perhaps sparse matrix
techniques would make its inversion go faster. 

Once the matrix is inverted the {\it solutions} go fast.  For the MR7
scheme, the solution time is about 0.6 seconds. This time scales as $NI
(2I+ \Delta i_{N-1})$, so if $N$ is kept constant it scales roughly as
$I^2$; thus, 4096 channels would take about 5 seconds.

\section{SCHEMAS FOR LO SETTINGS} \label{schemas}

In \S \ref{numbers} we found that the number of different LO
frequencies, $N$, must exceed 3. However, this tells us nothing about
how the quality of the solution is affected by $N$, and even less about
how the LO frequencies should be chosen, i.e.\ the values of $\Delta
i_n$.  It is not clear to this author how to investigate these matters
analytically. Rather, we turn to numerical experiments. Specifically, we
consider what we hope are intelligent schemas for $\Delta i_n$, adopt them
for a range of $N$, and analyze the results of the numerical
experiments.  We begin by discussing various schemas. First we describe
a conservative approach, which uniformly samples $\Delta i_{nn'}$ up to
a maximum but requires a fairly large number of $N$, and then we
describe some less conservative approaches.

\subsection{The Minimum-Redundancy (MR) Schema} \label{mra}

First we discuss Minimum-Redundancy (MR) settings having $N$ LO
settings. We will denote such settings by the symbol MRN, where N is the
number of LO settings. At the most conservative and basic level, common
sense suggests the following criteria:
\begin{enumerate}

\item An arithmetical progression for the successive frequency
differences $\Delta i_{nn'}$, which we also call ``spacings''. Spacings
should begin with unity and increment by unity up to some
maximum value $N_{max}$. It seems to us, intuitively, that incrementing
by unity is akin to sampling uniformly when doing Fourier transforms,
which is always the desirable situation.

\item A reasonably large value for the fractional LO coverage $h$, which
  is the maximum LO offset $\Delta i_{N-1}$ in units of the spectrometer
  bandwidth. It seems to us, intuitively, that precisely recovering
  broad-scale frequency structure requires sampling those broad scales
  with a comparably large value of $\Delta i_{N-1}$.

\end{enumerate}

	The problem of generating an arithmetic progression with $N$
settings of the LO frequency is akin, in radioastronomical
interferometry, to the well-studied problem of generating an arithmetic
progression of unidirectional baselines with a linear array of
telescopes. For $N$ LO frequencies there are $N(N-1) \over 2$ frequency
spacings, some of which are redundant; similarly, for $N$ telescopes on
the ground there are $N(N-1) \over 2$ distance spacings, some of which
are redundant. The classic discussion by Moffett (1968) considers these
minimum-redundancy telescope arrays, which use $N$ antennas to generate
a minimally redundant arithmetical progressive series for $\Delta
i_{nn'}$.  Zero redundancy is possible only for $N \leq 4$; for $N=4$
the spacings range up to $\Delta i_{N-1} = 6$. For $N > 4$ there must be
some redundancy. Moffett presents two types of minimum-redundancy
arrays, {\it restricted} and {\it general}.

	Restricted minimum-redundancy arrays provide all spacings
$\Delta i_{nn'}$ up to a maximum $N_{max}$, with no gaps; these are
useful for radio interferometry when the available real estate is
limited. General ones provide all spacings up to a particular limit, and
in addition provide larger spacings. For example, the $N=7$ restricted
array provides all spacings $\Delta i_{nn'} \leq 17$, while the $N=7$
general array provides all spacings $\Delta i_{nn'} \leq 18$ and, in
addition, $\Delta i_{nn'} = [24, 26, 31]$. For our purposes the general
array is a better choice: apart from the fact that we are not limited by
available real estate, it provides more different values for $\Delta
i_{nn'}$ and, also, a larger value of $h$. For the general
minimum-redundancy arrays, roughly $\Delta i_{N-1} \approx 31 \left({N
\over 7}\right)^{2.66}$. This is a steep dependence, so it is possible
to generate a large number of LO frequencies to get a desired $h$
without making $N$ ridiculously large.

\clearpage
{\tiny
\begin{deluxetable}{ccccccl} 
\tablewidth{0pc}
\tablecaption{Schemas: Definitions and Results \label{moffett} }
\tablehead{
\colhead{Schema} & \colhead{$\Delta RF$} & \colhead{$\sigma(IF)$} & \colhead{$F$-$Ampl[1]$} & \colhead{$ N_{max}$} & 
\colhead{$\Delta i_{N-1}$} & \colhead{LO spacings}}
\startdata
MR3 & --4.1 & 4.1 & 63.4 & 3 & 3 & $ \cdot 1 \cdot 2 \cdot $ \\ 
MR4 & --4.4 & 3.4 & 74.1 & 6 & 6 & $  \cdot 1 \cdot 3 \cdot 2 \cdot $ \\ 
MR5 & 0.149 & 1.19 & 25.9 & 9 & 13 & $ \cdot 4 \cdot 1 \cdot 2 \cdot 6 \cdot $ \\ 
MR6 & 0.051 & 1.27 & 16.8 & 13 & 19 & $  \cdot 6 \cdot 1 \cdot 2 \cdot 2 \cdot 8 \cdot $ \\ 
MR7 & 0.020 & 1.04 & 11.2 & 18 & 31 & $  \cdot 14 \cdot 1 \cdot 3 \cdot 6 \cdot 2 \cdot 5 \cdot $ \\ 
MR8 & 0.028 & 1.07 & 9.2 & 24 & 39 & $  \cdot 8 \cdot 10 \cdot 1 \cdot 3 \cdot 2 \cdot 7 \cdot 8 \cdot $
\\ 
MR9 & 0.005 &  1.04 & 9.4 & 29 & 29 & $  \cdot 1 \cdot 3 \cdot 6 \cdot 6 \cdot 6 \cdot 2 \cdot 3 \cdot 2
\cdot $ \\ 
MR10 &  0.003  & 0.93 & 5.2 & 37 & 73 & $  \cdot 16 \cdot 1 \cdot 11 \cdot 8 \cdot 6 \cdot 4 \cdot 3 \cdot
2 \cdot 22 \cdot $ \\ 
MR11 &  0.004 & 0.99 & 4.5 & 45 & 91 & $  \cdot 18 \cdot 1 \cdot 3 \cdot 9 \cdot 11 \cdot 6 \cdot 8 \cdot
2 \cdot 5  \cdot 28 \cdot $ \\ 
MR$3^2$    &  --4.3    & 2.20 & 61.6 & 1  & 5   & $  \cdot 1 \cdot  4 \cdot  $ \\
MR$4^2$    &  0.319    & 1.23 & 51.9 & 1  &  14  & $  \cdot 1  \cdot 9  \cdot 4 \cdot $ \\ 
MR$5^2$    &  0.012    & 1.04 & 5.8 & 1  & 57   & $  \cdot 16  \cdot 1 \cdot 4  \cdot 36 \cdot $ \\ 
MR$6^2$    &  0.013    & 1.08 & 3.1 & 1   & 109   & $ \cdot 36  \cdot 1  \cdot 4  \cdot 4  \cdot 64 \cdot $ \\ 
MR$3^{1.7}$   & --4.2  & 2.68 & 61.2 & 1 & 4   & $ \cdot 1  \cdot 3 \cdot  $ \\
MR$4^{1.7}$   & --2.6  & 1.64 & 106.3 & 1  &  11  & $ \cdot 1 \cdot  6 \cdot  4 \cdot $ \\ 
MR$5^{1.7}$   & 0.029  & 1.10 & 8.8 & 1  & 36   & $ \cdot 11 \cdot 1 \cdot 3 \cdot 21 \cdot $ \\ 
MR$6^{1.7}$   & 0.017  & 1.14 & 6.0 & 1  &  63  & $ \cdot 21 \cdot 1 \cdot 3 \cdot 4 \cdot 34 \cdot $ \\ 
$3^{1.7}$      & --4.1 & 4.17 & 62.0 &3  &  3  & $ \cdot 1  \cdot  2 \cdot  $ \\
$4^{1.7}$      &  0.451 & 1.89 &48.1 & 3  &  6  & $ \cdot 1 \cdot 2 \cdot 3 \cdot  $ \\
$5^{1.7}$      & 0.250 & 1.28 & 50.3 & 7  &  10  & $ \cdot 1 \cdot 2 \cdot 3 \cdot 4 \cdot  $ \\
$6^{1.7}$      & --2.7 & 1.68 & 129 & 11 & 15   & $ \cdot 1 \cdot 2  \cdot 3 \cdot 4 \cdot 4 \cdot  $ \\
$3^{\Delta 4}$ & 0.223 & 1.22 & 32.3 & 3  &  9  & $ \cdot  2 \cdot  1 \cdot  6 \cdot $ \\
$3^{\Delta 5}$ & 0.021 & 0.98 & 10.9 & 3  &  27  & $ \cdot 18 \cdot  2 \cdot  1 \cdot  6 \cdot $ \\
$3^{\Delta 6}$ & 0.021 & 1.17 & 4.2 & 3  &  81  & $ \cdot 18 \cdot  2 \cdot  1 \cdot  6 \cdot 54 \cdot $ \\
MR5,8-x & 0.019 & 1.64 & 25.7 & $8 \times 9$ & $8 \times 13$ & $8 \times [ \cdot 4 \cdot 1 \cdot 2 \cdot 6 \cdot ]$ \\ 
MR$5^2,8$-x &  0.011 & 1.30 & 5.6 & $8 \times 1$  & $8 \times 57$  & $8 \times[ \cdot 16  \cdot 1 \cdot 4  \cdot 36 \cdot] $ \\
MR7,2 &  0.068   & 1.21  & 5.4 & $2 \times 18$ & $2 \times 31$ & $ 2 \times [\cdot 14 \cdot 1 \cdot 3 \cdot 6 \cdot 2 \cdot 5 \cdot] $ \\ 
MR7,4 &  0.273   &  4.91 & 3.0 & $4 \times 18$ & $4 \times 31$ & $ 4 \times [ \cdot 14 \cdot 1 \cdot 3 \cdot 6 \cdot 2 \cdot 5 \cdot] $ \\ 
MR7,8 & $ 7 \times 10^6$ &  341 & 5.8 & $8 \times 18$ & $8 \times 31$ & $ 8 \times [  \cdot 14 \cdot 1 \cdot 3 \cdot 6 \cdot 2 \cdot 5 \cdot] $ \\ 
\enddata
{\tiny \tablecomments 
{Schemas are defined in the text, \S \ref{mra} and \S
  \ref{otherschemas}. Columns 2-4 contain our three quality indicators:
  the mean RF power offset and the RMS error for IF gains, defined in \S
  \ref{qualityindicators}, and the lowest-frequency Fourier amplitude,
  defined in \S \ref{relativefourier}. Spacings are completely covered
  up to $N_{max}$, which is given in column 5. Column 6 contains $\Delta
  i_{N-1}$, which is the maximum LO spacing and is equal to the sum of
  the $N-1$ spacings.}
}
\end{deluxetable}
}

The first nine rows of Table \ref{moffett} presents information about
the range $N=3\,$-11. Column 1 contains our names for the setting
arrangement, running from MR3 to MR11; columns 2-4 provide the three
quality indicators from the numerical experiments, which we define below
in \S \ref{numexpts}; columns 5-6 provide the maximum LO separation with
continuous coverage $N_{max}$ and the maximum LO separation $\Delta
i_{N-1}$. The last column, headed ``LO spacings'', is from the general
arrays in Moffett's (1968) Table 1. It has dots and numbers: the dots
represent the frequency settings of the LO and the numbers the spacing
between the frequencies in units of the minimum spacing. Ishiguro (1980)
discusses a subset of algorithms that generate arrays having larger $N$.

\subsection{Other Schemas} \label{otherschemas}

From the practical standpoint, an observer wants to keep $N$ as small as
possible. MRN settings obtain uniform coverage in $\Delta i_{nn'}$ but
require fairly large $N$ to attain large maximum separations $\Delta
i_{N-1}$. Here we propose schemas that sacrifice the uniformly-sampled
$\Delta i_{nn'}$ in favor of increasing the maximum separation $\Delta
i_{N-1}$.

We consider five such schemas, four of which are specified in terms of
the MRN spacings $\Delta i_{MRN}$. The five schemas are: \begin{enumerate}

\item The LO separations are the square of the MRN set, equal to $(\Delta
  i_{MRN})^2$. We designate this by MRN$^2$. This works quite well and
  we recommend it in our comparative discussion in \S \ref{bestlsfs}.

\item The LO separations are the 1.7-power of the MRN set, equal to
$(\Delta i_{MRN})^{1.7}$.  We designate this by MRN$^{1.7}$. This works
  almost as well as the MRN$^{2}$ schema.

\item The LO separations are equal to $-3^n$, where $n$ varies from 0 to
  $N-1$. For $1 \le n \le (N-1)$, the $n^{th}$ LO frequency is given by
  $\Delta i_n= \Delta i_{n-1} + (-3)^{n-1}$. We designate this by
  $3^{\Delta N}$. This works comparably to the MRN$^{1.7}$ schema. 

\item The LO {\it values} (not the separations) are a power-law
  series. The offset of the $n^{th}$ LO from the $0^{th}$ is equal to
  $n^{1.7}$, where $n$ varies from 0 to $N-1$. We designate this by
  N$^{1.7}$. This works less well than the above schemas.

\item The LO separations are equal to $2^{\Delta i_{MRN}}$. We designate this
  schema by $2^{MRN}$. This schema provides such poor results that we
  do not include it in Table \ref{moffett}. 

\end{enumerate}

\noindent The exponent 1.7 in schemas MRN$^{1.7}$ and N$^{1.7}$ is
inspired by the choice of spacings for the Very-Large-Array antennas. We
have not experimented with different values of the exponent.

\section{NUMERICAL EXPERIMENTS} \label{numexpts}

We evaluated the above schemas in numerical experiments. We first
invented a noise-free IF gain and RF power spectrum and, for each
schema, ran 256 trials. In each trial we added 2 K Gaussian-distributed
noise for each LO setting.

\begin{figure} [h!]
\begin{center}
\includegraphics[width=2.5in]{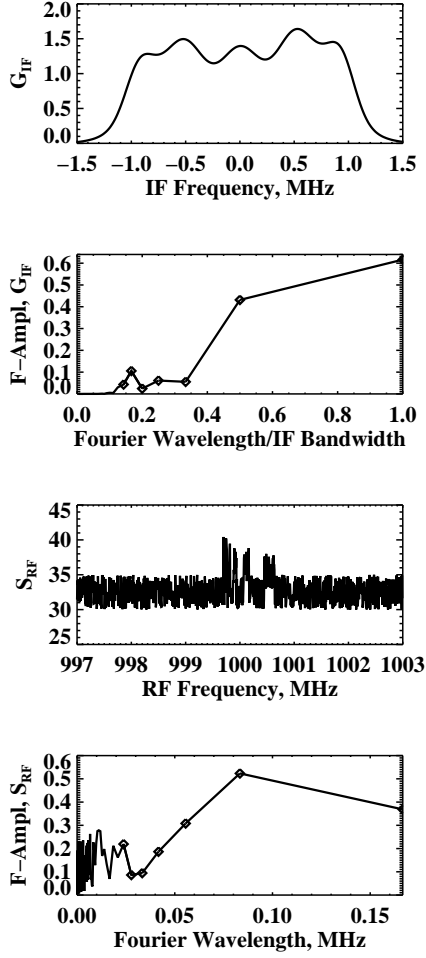} 
\end{center}
\caption{The noise-free input spectra for our numerical experiments. The
  top two panels display the IF gain spectrum versus IF frequency and
  its Fourier amplitude spectrum; the bottom two the RF power spectrum
  versus RF frequency and its Fourier amplitude
  spectrum. \label{plotfakedata}} 
\end{figure}

Figure \ref{plotfakedata} shows the noise-free input spectrum and its
Fourier amplitude spectrum for the IF gain (top two panels) and the RF
power (bottom two panels). The IF gain is the product of three terms, which
are meant to simulate three different hardware characteristics:
\begin{enumerate}

\begin{mathletters}
\item The overall shape of the band-limiting filter at the spectrometer
  input, 
\begin{equation}
G_{IF, filter} = 0.5 (\tanh[ 5(f_{IF} + 1)] -  \tanh[ 5(f_{IF} -
  1)] ) \ .
\end{equation}

\item A standing wave with period 0.5 MHz in the cable connecting the
  feed to the spectrometer

\begin{equation} \label{standingwave}
G_{IF, wave} = 1+ 0.1 \cos\left( 2\pi {f_{IF} \over 0.5} \right) \ .
\end{equation}

\item A slowly-varying polynomial-dependent gain from electronics and amplifiers

\begin{equation}
G_{IF, poly} = 1+ 0.1f_{IF} + 0.5 f_{IF}^2 \ .
\end{equation}
\end{mathletters}
\end{enumerate}

\noindent In all of the above, the IF frequency $f_{IF}$ runs from --1.5
to +1.5 MHz, which range is covered by $I=512$ channels. The top panel
of Figure \ref{plotfakedata} shows the product of these terms. The
second panel shows the Fourier amplitude spectrum, with the horizontal
axis being the Fourier component wavelength in units of the 3-MHz IF
bandpass. Thus, the standing wave term above in equation
\ref{standingwave} is clear in this plot: its period is 0.5 MHz, which
is $1 \over 6$ the 3-MHz IF bandwidth so it appears at 0.17 on the
horizontal axis. We could have labeled the horizontal axis ``Fourier
Wavelength, MHz'' and made the maximum $1 \over 3$ instead of 1.

\noindent Similarly, the RF power is the sum of three terms:
\begin{enumerate}

\item A frequency-independent system temperature of 30 K;

\item Five rectangular spectral lines of widths 2, 4, 8, 16, and 32
  channels having amplitudes 7, 6, 5, 4, 3 K and spaced so that they
  are nonoverlapping (we used rectangular lines to facilitate seeing
  degradation in frequency resolution);

\item Channel-to-channel 
uniformly-distributed random noise with amplitude limits 0 to 5 K. This
simulates a rich, crowded mm-wave spectrum containing a plethora of
molecular lines.
\end{enumerate}

\noindent The sum of all these components provides a channel-average
system temperature of slightly more than 32.5 K. The third panel of
Figure \ref{plotfakedata} shows the RF power over 6 MHz bandwidth. The
fourth panel shows the Fourier amplitude spectrum, with the horizontal
axis being the Fourier wavelength in units of MHz. That is, sinusoidal
ripples across the spectrum in panel 3 have a maximum period of 1 cycle
over the 6 MHz band. 

\subsection{Results: Two Quality Indicators} \label{qualityindicators}

\begin{figure} [h!]
\begin{center}
\includegraphics[width=3.0in]{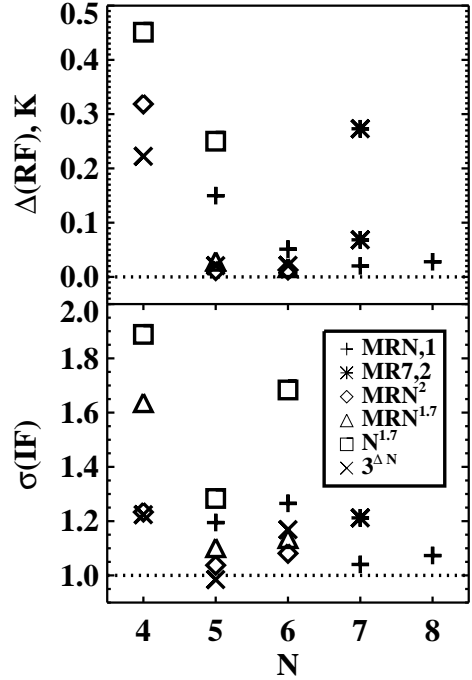} 
\end{center}
\caption{Plots of quality indicators $\Delta(RF)$ and $\sigma(IF)$
  versus the number of LO settings $N$. The different plot symbols
  specify the LO spacing schemas discussed in \S \ref{mra} and \S
  \ref{otherschemas}. The top panel displays the mean error of the RF power
  and the bottom the RMS uncertainty of the IF gain. \label{plt34}}
\end{figure}

Figure \ref{plt34} plots quality indicators for the above schemas as a
function of the number of LO settings $N$. Also, Table \ref{moffett}
lists these quantities in columns 2 and 3. We first consider two indicators:
\begin{enumerate}

\item The top panel displays $\Delta(RF)$, the mean error of the RF
power, where the mean is over two quantities, the channels in the RF
spectrum and the 256 trials. $\Delta(RF)$ represents an offset bias in
the derived RF powers. The units are Kelvins and the mean system
temperature is 32.5 K, so an error of 0.1 K is a fractional error of
0.3\%.

\item The bottom panel displays $\sigma(IF)$, the RMS uncertainty of the
  IF gain spectrum, in units of the theoretical RMS. To calculate this,
  we first form the difference between (a) the mean IF gain spectrum
  averaged over the 256 trials and (b) the input, noise-free IF gain
  spectrum. We consider only the central 200 channels and fit a
  second-order polynomial to eliminate broad baseline wiggles, which
  are more serious for smaller values of the fractional LO coverage
  $h$. With our 256 trials the number of independent measurements
  for each channel is $256 N$ and, with our 2 K input noise and 32.5 K
  system temperature, the ideal theoretical RMS is $2/32.5 \over \sqrt{
  256N}$.

\end{enumerate}

\noindent For completeness, we could also present the RMS uncertainty of
the RF power. However, those results are comparable to those of the IF
gain, so we refrain from presenting these to save space.

Both $\Delta(RF)$ and $\sigma(IF)$ decrease with $N$, and some schemas
are better than others. In particular, the MRN, MRN$^2$,MRN$^{1.7}$,
and $3^{\Delta N}$ schemas are all quite good.

\subsection{Results: Relative Fourier Amplitudes, a Third Quality
  Indicator} 
\label{relativefourier}

The above two quality indicators do not tell the whole story, for two
reasons. One, they are derived after baseline subtraction, which removes
large-scale ripples in the frequency spectra; we expect such ripples to
be larger when $h$, and thus $N$, is small. Two, schemas other than the
minimum-redundancy one do not uniformly sample Fourier components, and so
their Fourier amplitude spectra should be less uniform. Here the focus
is on the {\it relative} Fourier amplitudes, so we define all amplitude
spectra $F$-$Ampl$ to be the actual amplitude divided by the minimum
amplitude Fourier component for that spectrum.

\begin{figure} [h!]
\begin{center}
\includegraphics[width=4.0in]{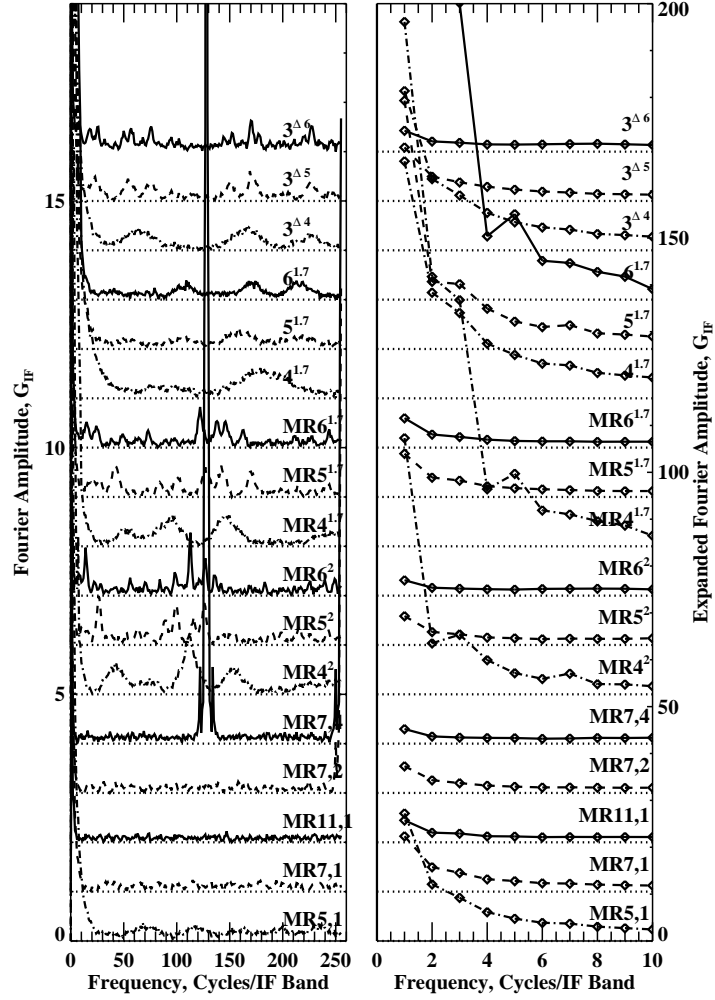} 
\end{center}
\caption{Plots of Fourier amplitude spectra of the IF gain residual
  spectra. The left panel exhibits the full frequency range from 1 cycle
  to 256 cycles per 512-channel IF band, with successive spectra
  displaced vertically by 1 unit. The right panel displays the low-frequency
  components by exhibiting the first 10 components (excluding the zero
  frequency component), with successive spectra displaced by
  10.625 units.
\label{plt36}} 
\end{figure}

Figure \ref{plt36} displays Fourier amplitudes of the IF gain residual
spectra for a selection of our schemas. The left-hand panel emphasizes
the higher frequencies.  The bottom three plots are for MR5,1, MR7,1,
and MR11,1. With the uniform sampling in $\Delta i_{nn'}$, these provide
quite uniform Fourier amplitudes at high frequencies; and as $N$
increases from 5 to 11, the fractional LO coverage $h$ increases, which
decreases the amplitude of lower-frequency Fourier components. For MR7,2
and MR7,4, we have strong Fourier components at frequencies 256 and 128
cycles, respectively, over the 512-channel IF band. These Fourier
components are easily understood, because they correspond to periods of
2 and 4 channels, respectively. The other schemas do not have uniform
sampling in $\Delta i_{nn'}$, and this is reflected in their Fourier
amplitudes, which are not very uniform at high frequencies.

The right-hand panel emphasizes the low frequencies. Generally, the
low-frequency amplitudes are smaller for large fractional LO
coverage $h$. This is reflected in the MR7,1, MR7,2, and MR7,4 spectra,
as well as the MR7,1 and MR11,1 spectra. Also the other schemas
emphasize larger $h$ at the expense of uniform coverage in $\Delta
i_{nn'}$, and this is reflected in their smaller low-frequency
amplitudes.

\begin{figure} [h!]
\begin{center}
\includegraphics[width=3.0in]{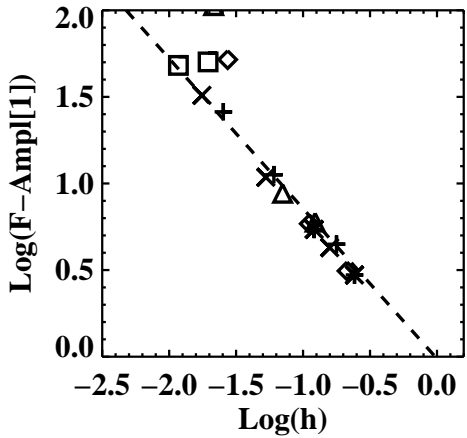} 
\end{center}
\caption{The amplitude of the first Fourier coefficient versus $h$, the
  fractional LO coverage. The dashed line is
  minimum-absolute-residual-sum fit to the points.  \label{plt37}}
\end{figure}

The first Fourier coefficient serves as a proxy for the low-frequency
Fourier components. Figure \ref{plt37} plots the first Fourier amplitude
$F$-$Ampl[1]$ versus the
fractional LO coverage $h$. As we anticipated, $F$-$Ampl[1]$ decreases with
$h$; the dashed line is a
minimum-absolute-residual-sum fit to the points, which fit de-emphasizes
large deviations from the fit. The fit is

\begin{equation} \label{fractional}
F{\rm -}Ampl[1] \approx h^{-0.87} \ .
\end{equation}

\noindent This parameter, $F$-$Ampl[1]$, is our third quality
indicator, and we list it in column 4 of Table \ref{moffett}. For all
reasonable schemas it is closely approximated by equation
\ref{fractional}.

\subsection{Summary: Which LSFS Schemas Are Best?} \label{bestlsfs}

Which LSFS schema is best? If one wants the best accuracy and large $N$
is not a problem, then the Minimum Redundancy (MRN) schema is ideal
because it provides a flat, featureless spectrum of high-frequency
Fourier amplitudes in Figure \ref{plt36}. If one is willing to sacrifice
accuracy in favor of a smaller number of LO settings $N$, then one can
consider the $MRN^2$, $MRN^{1.7}$, and $3^{\Delta N}$ schemas. These
provide comparable results for $N=5$ and $N=6$, with $MRN^2$ having the
edge. As a compromise between practicality and good results, our
numerical experiments suggest the $MRN^2$ schema, with $N \ge 4$.

\section{SCHEMAS THAT HAVE min($\Delta i_{nn'}) > 1$} \label{otherschemas1}

It is not strictly necessary for $\min (\Delta i_{nn'})$ to equal unity;
rather, it can equal some integer multiple, which we call
$R$. Increasing $R$ provides increased fractional LO coverage $h$, which
may be desirable. We envision two circumstances where $R>1$ might be
desirable.  One is when spectral resolution can be degraded and one can
bin the data into groups of $R$ channels (\S \ref{binning}). The other
is when the number of spectral channels $I$ is large: matrix sizes scale
with $I$ and the computational load for inverting a matrix scales with
the cube of the matrix size, so using LSFS with large $I$ and $R=1$ can
be a problem.  We can reduce this problem by using $R^{th}$ sampling (\S
\ref{rsampling}). 

\subsection{Binning} \label{binning}

If one wants to derive the IF gain spectrum and has prior knowledge that
it varies slowly with frequency, then one can increase the fractional LO
coverage $h$ by increasing $R$ and, also, bin the data into $R$ channels
and sacrifice resolution. Mathematically, this combination is identical
to a dataset with $R=1$.

\subsection{$R^{th}$ Sampling} \label{rsampling}

This technique retains full spectral resolution while keeping the matrix
sizes manageable by using the following subterfuge.  Suppose
$I=4096$. Convert the 4096-long spectrum into a series of $R$
subspectra, each of length $4096 \over R$, by choosing every $R^{th}$
point. For example, for $R=8$, one creates 8 subspectra, each 512 channels
long, each spanning nearly the full frequency range covered by the
original 4096 channels. One uses LSFS on the $R$ subspectra independently
and patches the $R$ result spectra together. For each subspectrum, the
smallest LO spacing is equal to the bin separation, so each solution is
mathematically identical to that for 512 channels and $R=1$. 

We did our numerical experiment for two schemes, the MR5,8-x scheme and the
MR5$^2$,8-x scheme; here the suffix ``-x'' signifies this $R^{th}$
sampling scheme.  Table \ref{moffett} lists the quality indicators for
these two experiments; they are comparable to the $R=1$ versions, which
is reasonable because the mathematical equivalence of the $R$ solutions.

\begin{figure} [p!]
\begin{center}
\includegraphics[width=5in]{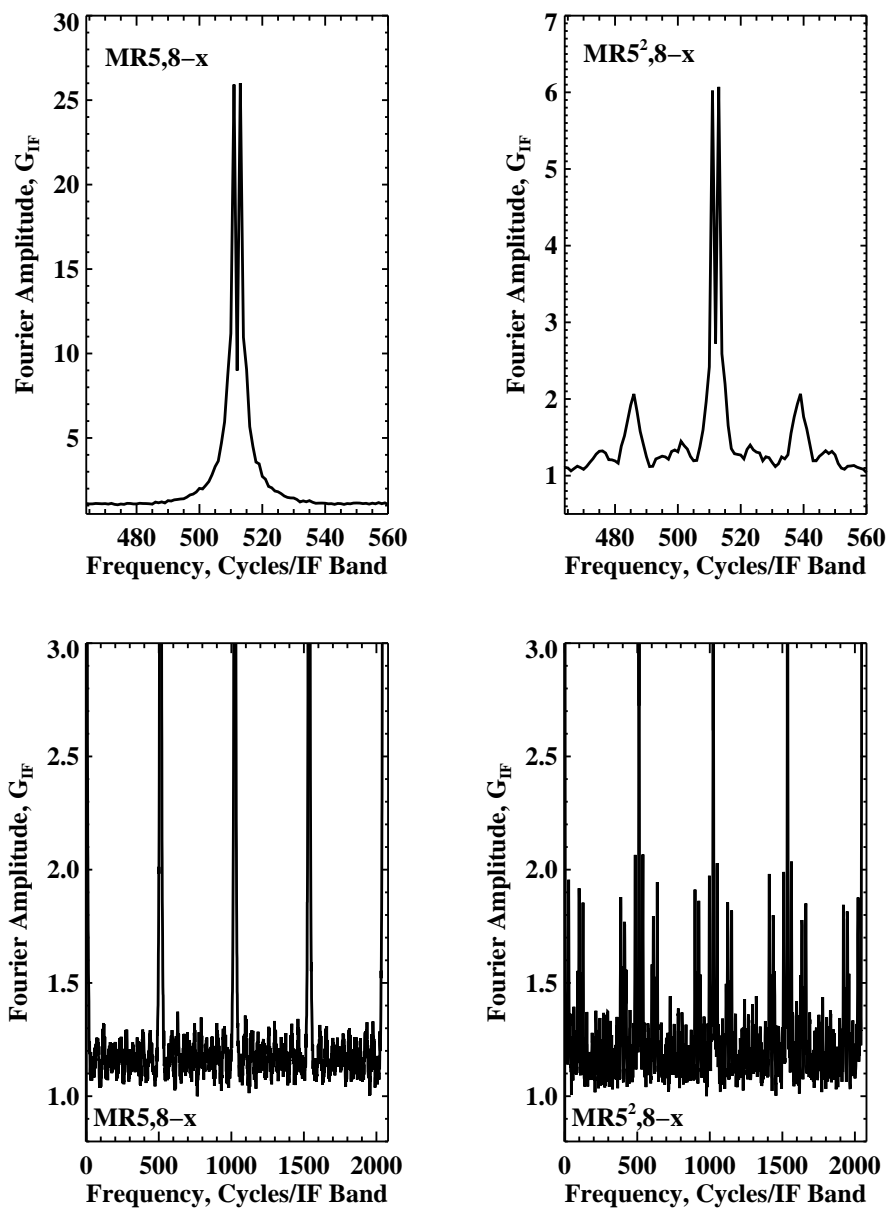} 
\end{center}
\caption{Numerical experimental Fourier amplitude spectra of the IF gain
  spectra for the two schemes MR5,8-x (left panels) and MR5$^2$,8-x (right
  panels). The bottom panels show the full frequency range and cut off
  the peaks; the top panels show a magnified frequency range centered at
  512 cycles per IF band. Note the difference in vertical scale for the
  top panels! The peaks repeat almost identically at multiples of 512.
  \label{plt368k}}
\end{figure}

The method is not perfect. Figure \ref{plt368k} shows the Fourier
amplitude spectrum of the 4096-channel IF gain spectrum. The bottom
panels show the full range of frequencies, 1 to 2048 cycles over the
4096-channel IF band. The most striking thing about these spectra are
the spikes, which lie at multiplies of 512 cycles per IF band. This is
easily understood, because the band contains 4096 channels; 512 cycles
per IF band corresponds to a period of 8 channels, which is the value of
$R$. This shows that each of the $R$ subspectra is slightly offset in
power from the others, which is a result of their being reduced
independently. Next, spikes repeat, almost identically, at multiples of
512. This repetition can be understood in terms of the Fourier
convolution theorem: the original data are, in effect, multiplied by a
function that repeats every $R$ channels; in the transform domain, the
spectrum is convolved by the Fourier transform of this function, which
produces the repetition.

The left-hand panels of Figurer \ref{plt368k} show MRN,R-x and the right
MRN$^2$,R-x. The top panels show expanded views around 512 cycles per IF
band. Note the difference in scale for the
top panels: the MRN,R-x schema has much bigger Fourier peaks at the
512-cycle multiples and much smaller and more uniform Fourier amplitudes
in between. In contrast, the MRN$^2$,R-x has about four times smaller
peaks, but stronger and less uniform Fourier amplitudes in between.

One could reduce these Fourier spikes and gain accuracy in the derived
results by associating an additive constant for each of the $R$ result
spectra and devising a minimization procedure to determine the values of
the $R$ constants. More appealing is using the conventional, well-known
technique of Wiener filtering to eliminate the spikes. Wiener filtering
would work especially well for the MRN,R-x result because the Fourier
power is so concentrated.

\subsection{No Binning: the MR7,R Schemas} \label{nobinning}

\begin{figure} [h!]
\begin{center}
\includegraphics[width=4.0in]{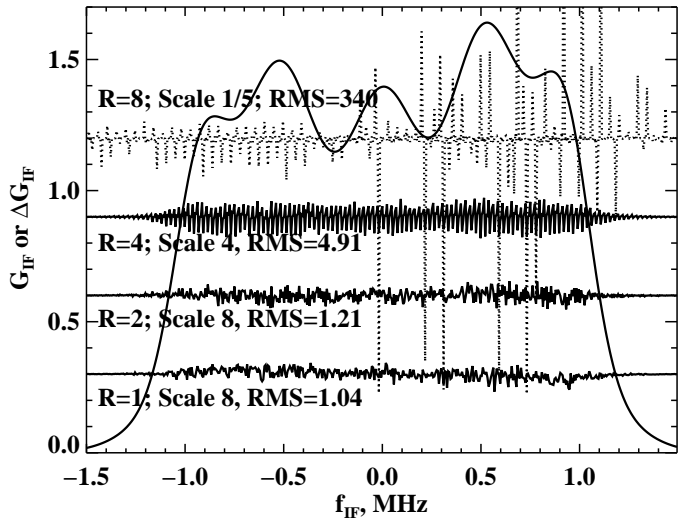} 
\end{center}
\caption{The noisy lines are the IF-gain difference spectra
  for $R=[1,2,4,8]$  versus $f_{IF}$ for the MR7,R schema discussed in
  \S \ref{nobinning}, scaled by the factors
  shown. The solid smooth line is the IF gain $G_{IF}$ itself.
\label{plt10}} 
\end{figure}

Suppose that one observes with $R \neq 1$ but does {\it neither} of the
above tricks---neither binning nor $R^{th}$ sampling. How do the
solutions fare under these conditions? We performed numerical
experiments for $R=[2,4,8,16]$ only for the MR7 scheme, and we denote
these LO setting arrangements by the name MR7,R.  We present the three
quality indicators in Table \ref{moffett} and show the derived IF gain
spectra in Figure \ref{plt10}.  As expected, larger $R$ leads to smaller
$F$-$Ampl[1]$, but the other two quality indicators are degraded. $R\geq
8$ breaks down and is worthless.

Figure \ref{plt10} exhibits IF gain
difference spectra for $R=[1,2,4,8]$, plotted against the background of
the IF gain itself. The results for $R=1$ and 2 look reasonable. The
result for $R=4$ looks reasonable except for the systematic periodic
signal in the difference spectrum. This has a period of exactly 4
channels, which is the minimum LO spacing, and confirms our hunch that
one cannot accurately reconstruct the input signals with Fourier
components smaller than the minimum LO spacing. Of course, The Fourier
amplitude spectra in Figure
\ref{plt36} show this behavior quantitatively, and shows also that for
$R=2$ there is also a systematic periodicity of 2 channels.  For $R=8$
the solution breaks down and becomes worthless; moreover, for 16 of the
256 trials the iterative solution discussed in \S \ref{iterative} didn't
converge. For $R=16$ the solutions never converged.

These results show that, with $R>1$, one should use either the binning
or the $R^{th}$ sampling trick. 

\section{POLARIZED STOKES PARAMETERS: SWITCHING AND LSFS}
\label{stokes}

Our above discussion applies to a single polarization, or to Stokes
$I$. Here we generalize to the three polarized Stokes parameters
$(Q,U,V)$.  For discussion purposes, we discuss the example of native
orthogonal linear polarizations, which produce time-variable voltages
$X$ and $Y$. We form Stokes parameters by taking time averages of the
four possible products, whose results we denote by, for example,
$XX$. The four Stokes parameters include the sum $(XX + YY)$, the
difference $(XX - YY)$, the product $2XY$, and the product with
$90^\circ$ phase lag $2YX$; for native linear polarizations, these
combinations produce Stokes $I$, $Q$, $U$, and $V$, respectively. (For
native circular polarizations, these combinations would produce Stokes
$I$, $V$, $U$, and $Q$, respectively.) Part of the process of
calibrating the Stokes parameters involves applying the Mueller matrix
calibrations to these combinations as discussed by Heiles et al.\
(2001), and we assume these corrections have applied to these
time-average products before applying the discussion below, which covers
how to obtain Stokes spectra from switching or the LSFS technique.

\subsection{Position and Frequency Switching}

Generating $I$ and $Q$ requires taking the sums and differences. For
position and frequency switching, this is simply a matter of
arithmetically combining the results in equations \ref{posswitch} and
\ref{freqswitch}, respectively.

Generating $U$ and $V$ requires crosscorrelating the
polarizations. Consider $U$, which we obtain from $2XY$. We write the
counterpart of equation  \ref{basic} as 

\begin{equation} \label{eqnbasicequiv}
P_U(f_{IF}) = [G_{XX,IF}(f_{IF}) \ G_{XX,RF}(f_{RF})]^{1/2} \ 
              [G_{YY,IF}(f_{IF}) \ G_{YY,RF}(f_{RF})]^{1/2} \ 
\end{equation}
$$        \left[ (U_A(f_{RF}) + U_A) + (U_R(f_{RF}) + U_R) \right] \ . $$

\noindent Here the square-root of the gain products $G$ appear because
$U$ is derived by multiplying voltages $X$ and $Y$, while the gains
$G_{XX}$ and $G_{YY}$ are power gains. We could write the equivalent of
equations \ref{posswitch} and \ref{freqswitch}, but these would be
algebraicly cumbersome and would not convey much information. Usually,
astronomical polarized signals are weak, so if we keep only zeroth order
terms then the position- and frequency-switched spectra are simplified
equivalents of equations \ref{posswitch} and \ref{freqswitch}, and they
both become

\begin{equation} 
{P_U(f_{IF}) -P_U'(f_{IF}) \over [ P_{XX}'(f_{IF})P_{YY}'(f_{IF})]^{1/2}} \approx 
	[U_A(f_{RF}) + (U_A - U'_A)] 
 \ .
\end{equation}

\subsection{LSFS}

	LSFS also applies to polarized Stokes parameters obtained from
crosscorrelation.  The application of the LSFS procedure to $XX$ and
$YY$ individually provides their associated RF powers and, more
importantly for now, the IF gains $G_{XX,IF}(f_{IF})$ and
$G_{YY,IF}(f_{IF})$.  Thus, in equation \ref{eqnbasicequiv}, we can
treat them as known quantities and move them to the left-hand side,
which is the same as pre-correcting the data for the IF gains.

	With this, it is straightforward to duplicate the steps leading
to equation \ref{lsone}.  We begin with the analog of equation
\ref{define_s},
\begin{equation}
S_U(f_{RF}) = [G_{XX,RF}(f_{RF}) \ G_{YY,RF}(f_{RF})]^{1/2} \ 
\left[ (U_A(f_{RF}) + U_A) + (U_R(f_{RF}) + U_R) \right] \ ,
\end{equation}

\noindent and carrying through the algebra we arrive at the equivalent of
equation \ref{lsone},

\begin{equation} \label{basicyx}
{P_{U, \; i,\Delta i_n} \over G_{IF, \; i}} = S_{U, \; i + \Delta i_n} \ ,
\end{equation}

\noindent where to reduce clutter we write $G_{IF, \; i} =
[G_{XX,IF, \; i} \ G_{YY,IF, \; i}]^{1/2}$.

The left-hand side contains the measured quantities and the right the
desired unknown ones.  In contrast to equation \ref{lsone}, there is
only a single unknown on the right-hand side. This means that the
least-squares solution for the unknowns is just the appropriate average
of the left-hand quantities.  Alternatively, we can follow the line of
development pursued in \S \ref{lsfs0} and express the solution as a
least-squares problem using matrices.  This is more time-consuming
computationally, but much
simpler programmatically and offers more flexibility.  In contrast to the
the situation in \S \ref{lsfs0}, this least-squares fit is a linear one
so it is not necessary to do an iterative solution. Neither do we need to
add the additional constraint embodied in equation
\ref{powerconservation}.

	Referring to \S \ref{numbers}, here the number of measurements
is $M = NI$ ($I$ channels for each of the $N$ LO frequencies) and
because we do not have to determine the gains ($\delta G_i$) the number
of unknowns is only $a = (I + \Delta i_{N-1})$.  To make the number of
measurements exceed the number of unknowns, we require $N > (1+h)$,
which is more easily satisfied than the corresponding equation
\ref{nvsh}.  The current $\bf X$ matrix has $NI$ rows and $(I+\Delta
i_{N-1})$ columns. For our textbook example of \S \ref{textbook}, 
the correspondent to equation \ref{xmatrixeqn}
looks like

\begin{eqnarray} \label{xmatrixeqn1}
{\bf X} = \left[
\begin{array} {ccccccccccc}
1 & 0 & 0 & 0 & 0 & 0 & 0 \\
0 & 1 & 0 & 0 & 0 & 0 & 0 \\
0 & 0 & 1 & 0 & 0 & 0 & 0 \\
0 & 0 & 0 & 1 & 0 & 0 & 0 \\
0 & 1 & 0 & 0 & 0 & 0 & 0 \\
0 & 0 & 1 & 0 & 0 & 0 & 0 \\
0 & 0 & 0 & 1 & 0 & 0 & 0 \\
0 & 0 & 0 & 0 & 1 & 0 & 0 \\
0 & 0 & 0 & 1 & 0 & 0 & 0 \\
0 & 0 & 0 & 0 & 1 & 0 & 0 \\
0 & 0 & 0 & 0 & 0 & 1 & 0 \\
0 & 0 & 0 & 0 & 0 & 0 & 1 \\
\end{array}
\right] \ .
\end{eqnarray}

\noindent In this matrix, the arrangement of rows and columns is similar
to that in equation \ref{xmatrixeqn}, except that the IF gains do not
appear. Here in equation \ref{xmatrixeqn1}, we have \begin{enumerate}

	\item The first four rows pertain to the lowest LO frequency with
$\Delta i_0 = 0$

	\item The next two pairs of four rows pertain to $\Delta i_1=1$
and $\Delta i_2= 3$.

	\item The last seven columns are the coefficients of the seven
	RF powers $\delta s_{i + \Delta i_n}$ in equation \ref{final_ls}.
\end{enumerate}

\section{LSFS versus Switching} \label{comparison}

The traditional technique for heterodyne spectroscopy is called
``switching''. It involves taking an off-source reference spectrum. This
technique has been used by radioastronomers for decades. We offer a few
comments that compare the two techniques. These comments are based
mainly on our observing experience over the years and the numerical
simulations discussed above. Unfortunately, we have not experimentally
investigated these matters with LSFS spectra because our observing
experience with this technique is limited to a handful of projects.

\subsection{Channel-to-channel noise} 

First, consider $\sigma (IF)$, the channel-to-channel noise for the
IF gain. Figure \ref{plt34} displays $\sigma (IF)$ for various LSFS
schemas. These dispersions are normalized to the ideal, for which the
noise is determined by the time-bandwidth product---where the time is
the {\it full} integration time used for all LO settings. A value of
unity would result if the LSFS fit provided no degradation in noise. All
the values shown on Figure \ref{plt34} are less than 2.

It is important to recall the noise in a conventional position- or
frequency-switched spectrum. Conventionally, for a switched spectrum half
the total observing time is spent on the OFF spectrum; this
reduces the sensitivity of the ON spectrum by $\sqrt 2$. Subtracting the equally
noisy OFF spectrum reduces the sensitivity by another factor $\sqrt
2$. Thus, a conventional switched spectrum has $\sigma(IF)=2$. All of
the points displayed on Figure \ref{plt34} have better sensitivity than
a conventional switched spectrum!

\subsection{Baseline wiggles} \label{baselinewiggles}

Next, consider the slow undulations of the RF power spectrum with
frequency, commonly referred to as ``baseline wiggles''. For
conventional switched spectra, our discussion of \S \ref{review} shows
that the baseline wiggles are determined by the frequency dependence of
the RF power, which is in turn determined by reflections, the RF gain,
and system temperature. These are instrumental properties associated
with the telescope structure, RF amplifiers, and associated circuitry;
they change slowly with time. Thus they do not tend to decrease with
increased integration time.

For LSFS spectra, there are two sources of baseline wiggles. One is
these {\it intrinsic} wiggles that reside in the RF power
spectrum. These often occur from reflections, as explained in \S
\ref{introduction}. LSFS will not eliminate these.  On the contrary, it
is very useful for determining them, just as we did in our detailed
discussion of reflection-induced baseline wiggles at
Arecibo\footnote{See Heiles, Carl 2005, ``Some Characteristics of ALFA's
  Fixed Pattern Noise (FPN)'', Arecibo Technical memo 2005-04, available
  at {\tt http://www.naic.edu/science/techmemos\_set.htm}}.  We
emphasize that these intrinsic wiggles are not artifacts introduced by
LSFS because they are actually present in the RF power spectrum that
enters the feed. LSFS will determine these intrinsic wiggles but, in
contrast to frequency-switched spectra, it will not exacerbate them.

The other type of baseline wiggle in LSFS spectra is associated with
the {\it fitting} of low-frequency Fourier amplitudes shown in the
right-hand panel of Figure \ref{plt36}. The nonflat Fourier amplitudes
reveal that some Fourier components are reproduced less accurately than
others. In particular, the amplitudes increase for larger wavelengths,
i.e., for slower variations with frequency---leading to baseline
wiggles.

This fitting type of LSFS wiggle should decrease with increased
integration time because there should be no systematic bias in the {\it
phase} of the fitted LSFS Fourier components. Rather, the amplitudes are
less well determined, leading to increased noise, but the location of a
positive-going fitted baseline ripple should change from one integration
to the next.

\section{SUMMARY} \label{summary}

We described a new technique for obtaining accurate results from
heterodyne spectroscopy. It involves taking measurements at 3 or more
local oscillator (LO) frequencies and using least squares to derive the
RF power and IF gain spectra as individual entities. We call this the
Least-Squares Frequency-Switching (LSFS) technique.  We have used the
technique in two ways: one, to obtain a single IF gain spectrum used to
correct a series of several thousand mapping measurements; and two, to
obtain the RF power spectrum directly during a long integration.

We discussed mathematical details and computational requirements of the
technique and explored optimum observing schemas using numerical
experiments. The quality of the results depends on the choice of the LO
frequencies, and \S \ref{bestlsfs} summarizes our results and
recommendations.

We illustrate the method with three real-life examples. In one of
practical interest (\S \ref{realworld}), we created a large contiguous
bandwidth aligning three smaller bandwidths end-to-end; radio
astronomers are often faced with the need for a larger contiguous
bandwidth than is provided with the available correlator.  In \S
\ref{otherschemas1} we outlined an approach suitable for computationally
difficult cases as the number of spectral channels grows beyond several
thousand.

\acknowledgements This work was supported in part by NSF grant AST-0406987
and, also, by the NAIC.  Josh Goldston Peek and Snezana Stanimirovi{\'c}
read an early version of the manuscript and made several important
suggestions, and Benjamin Winkel made several valuable comments on a
later draft.

\clearpage
\clearpage

\clearpage

\clearpage

\clearpage

\clearpage

\clearpage

\clearpage

\clearpage

\end{document}